\documentclass[runningheads]{llncs}
\usepackage[T1]{fontenc}
\usepackage{graphicx}
\usepackage[svgnames]{xcolor}
\usepackage{multirow}
\usepackage{todonotes}
\usepackage{hyperref}
\usepackage{booktabs}
\usepackage{graphicx}%
\usepackage{multirow}%
\usepackage{textcomp}
\usepackage{pdfpages}

\begin{document}
\title{Multi-Platform Framing Analysis: A Case Study of Kristiansand Quran Burning}
%

\author{Anna-Katharina Jung\inst{1}
\orcidID{0000-0002-0905-4932} 
\and
Gautam Kishore Shahi\inst{1}
\orcidID{0000-0001-6168-0132} \and
Jennifer Fromm\inst{1}
\and
Kari Anne Røysland\inst{2} 
\and \\
Kim Henrik Gronert\inst{3}
}
\authorrunning{Jung et al.}
%

\institute{University of Duisburg-Essen, Duisburg, Germany  \\ \email{gautam.shahi@uni-due.de}\and
University of Agder, Kristiansand, Norway 
\and
Kristiansand Kommune, Kristiansand, Norway\\
}

\maketitle
\begin{abstract}

The framing of events in various media and discourse spaces is crucial in the era of misinformation and polarization. Many studies, however, are limited to specific media or networks, disregarding the importance of cross-platform diffusion. This study overcomes that limitation by conducting a multi-platform framing analysis on Twitter, YouTube, and traditional media analyzing the 2019 Koran burning in Kristiansand, Norway. It examines media and policy frames and uncovers network connections through shared URLs. The findings show that online news emphasizes the incident's legality, while social media focuses on its morality, with harsh hate speech prevalent in YouTube comments. Additionally, YouTube is identified as the most self-contained community, whereas Twitter is the most open to external inputs.

\end{abstract}

\section{Introduction}
\label{sec:1}

The appearance and rise of social media platforms restructured and diversified the process of information diffusion. While priorly, the dissemination of information was limited to traditional media outlets managed by gatekeeping journalists, nowadays, information can be produced and shared by everyone with online access \cite{bruns2018gatewatching}. As a result of this development, journalists have not only lost their major influence on the dissemination of content but also the sovereignty of interpretation. The classification of content is in the hands of every single online actor.The process of highlighting certain elements of a piece of content and promoting a particular understanding and interpretation is known as the concept of framing \cite{entman1993framing}. Although the media landscape is diverse, previous research on framing and information diffusion often focused only on one specific social media platform or traditional news outlets  \cite{jung2018information}. With our study, we aim to advance multi-platform framing of societal incidents. The case of the Quran burning in Kristiansand (Norway) is used as an example in our study to demonstrate the application and diffusion of media frames throughout different types of online media platforms. On 16th November 2019, Lars Thorsen - the leader of the Norwegian Anti-Islam group Stop the Islamification of Norway(SIAN), attempted to burn the Quran on the main square of Kristiansand. Several persons attacked Lars Thorsen to stop him from burning the Quran and were arrested by the police. About 300 people witnessed the incident, and shortly afterwards, videos of the Quran burning and the attacks circulated on the net. The incident was heavily discussed in both online news sources and social media. While many members of the Muslim community described the attackers as defenders of Islam, other actors rather expressed anti-Islamist views. These different media frames resulted in tensions within Norwegian society and considerable problems for the Norwegian government. We aim to answer the following research questions: \\
RQ1: How was the Quran burning incident in Kristiansand framed on different social media platforms (Twitter (now X), YouTube) and on online news sites?\\
RQ2: How is the diffusion between the different social media platforms and online news sites shaped?\\
For this multi-platform framing analysis, we analyzed 1,136 tweets, 71 YouTube videos with 2031 comments, and 109 articles from online news sources. We distinguish between social media that allows any user to create and share content and online news sites where editorial teams retain control over the publication of articles and associated comments \cite{hasan2009s}. It should be noted that online news sites differ in the extent to which they follow the ethical code of practice for the press adopted by the Norwegian Press Association. The study is organized as follows. First, a short insight into the development and methodical approaches to the framing theory are presented, followed by a concise overview of multi-platform framing and cross-platform diffusion. Afterwards, the methodical approach is presented. Summarizing the case study, the approach to data collection and cleaning, such as the description of the coding procedure and code book. This is followed by the presentation of the results, giving an overview about the distribution of frames on the different platforms, such as the URL analysis. In the discussion section, connections to the state of the art presented and prior research are drawn. This discussion section is followed by a short overview about the limitations of the study and the ideas for future research. The study is rounded off by the presentation of the conclusion in which the main findings and the answers to the research questions are summarized.


\section{State of the Art}
\label{sec:2}

In this section, we describe framing theory, multi-platform framing and Cross platform information diffusion.

\subsection{Framing theory}
The theory of framing was originally created in the field of sociology in the 1950s and has been refined ever since by various disciplines ranging from psychology to political and media studies \cite{bateson1972theory,goffman1974frame,schaffner2010winning}. As well in the more technical field of information systems, the framing theory has been adapted to approach the analysis of stakeholder perspectives in a technological context, or to analyze communication in online environments  \cite{orlikowski1994technological,jung2018information} One of the most comprehensive definitions of framing was provided by the political scientist Robert Entman: “Framing essentially involves selection and salience. To frame is to select some aspects of a perceived reality and make them more salient in a communication text, in such a way as to promote a particular problem definition, causal interpretation, moral evaluation and/or treatment recommendation for the item described” ((\cite{entman1993framing}, p. 52). Thus, framing looks not only at the issues and topics covered by the media but also at the angles being taken. As diverse as the disciplines that adapted the framing theory are the methodological approaches that have been developed for its analysis. The operationalization of frames for empirical studies is complex and challenging, among other things, due to the huge amount of data that needs to be annotated (manually) and the influence of personal interpretation of media and its frames \cite{touri2015using}. In prior research, authors divide the existing methodological approaches to frame analysis in deductive approaches (based on priorly formulated frames) and inductive approaches (based on frames derived from the specific data set in focus) \cite{touri2015using}. While studies solely based on deductive frame categories are very systematic and can be easily replicated, they have the shortcoming that they might ignore important information within the data set if it does not fit the priorly defined frames \cite{matthes2008content}. In contrast to that, inductive approaches use the data and its context as a basis for the creation of frames and thus are much more sensitive to the peculiarities of specific research cases \cite{van2012frames}. In our study, we applied a mixture of deductive and inductive approaches to the framing analysis. As a basis, we used the codebook for analysis of media frames within and across policy issues by \cite{boydstun2014tracking,card2015media} 
and modified it according to the cultural and thematic context. The codebook will be explained in more detail in the method section.

\subsection{Multi-platform framing}

While journalists and traditional media houses still have a huge impact on the shaping of public debates, the influence of online users not bound to news values and reporting standards has increased with the rise of social networks and hybrid media systems \cite{chadwick2015politics}. This development is described as networked gatekeeping and networked framing and was defined by \cite{meraz2013networked} as the involvement of a diversity of online actors from many different backgrounds, including journalists, activists, non-elite media supporters and regular users, who gained prominence and attention in their network by effective communicative and social practices to spread their messages. Networked framing, like networked gatekeeping, stands for the prominence interpretations received via crowdsourcing actions \cite{meraz2013networked}. Although \cite{meraz2013networked} value the difference between gatekeeping and framing in social networks within their own research on the revolution in Egypt, they fully focus on the discourse on Twitter and thus miss out on giving an overview of the process on multiple platforms. As well studies incorporating the idea of networked publics and framing mainly limited their scope to one platform, as for example \cite{kermani2021mapping} who mapped the discussion on the Iranian presidential elections 2017 on Twitter \cite{kermani2021mapping}, depicting the different groups of online elites discussing the subject or previous research \cite{walsh2023digital} which analyzed the discourse on migration during the Canadian elections 2019. Another recent example is the analysis of \cite{knupfer2022hijacking} who analyzed the hijacking and reframing of the MeToo hashtag by right-wing actors in a multi-national Twitter analysis. While \cite{knupfer2022hijacking} do involve different languages and thus different national communities on Twitter, they did disregard other media outlets than Twitter. One of the very few studies which pays attention to the research gap of multi-platform framing is \cite{poyhtari2021refugee}, which analyzed the discourse about the refugee crisis in the Finish online news media and social media \cite{poyhtari2021refugee}. In contrast to the present study, a computational topic and framing detection were applied, and a latent Dirichlet allocation algorithm was used. While the scope of our study is similar, the methodical approaches differ from each other. 

\subsection{Cross platform information diffusion}
The sharing and presentation of frames within one network and across platform boarders can be as well described as a form of information diffusion. Information diffusion can be defined “as the process by which a piece of information (knowledge) is spread and reaches individuals through interactions” \cite{zafarani2014social}. On social media, this process depends on individuals who spread information through retweets, shares and likes. The information created by verified accounts spreads faster than non-verified sources \cite{shahi2021exploratory,shahi2022mitigating}. One stream of information diffusion research takes a micro perspective and aims to understand why individual users distribute information \cite{liu2020homogeneity,syn2015social}. For example, authors in prior research \cite{syn2015social} found that learning and social engagement are the most important motivations for content sharing \cite{syn2015social}. Another stream of research examines the phenomenon from a macro perspective, focusing on predicting how information spreads through a social network. These studies often aim to assess and improve information diffusion models such as information cascade or threshold models \cite{su2018toward,sela2018active,hosseini2019assessing,molaei2020deep}. To understand information diffusion, scholars highlighted the importance of analyzing the interplay of different actor and content characteristics \cite{han2020importance}. Actor-centered studies demonstrated the power of opinion leaders in the information diffusion process \cite{probst2013will,wang2020understanding} and distinguish between different roles such as information starters, information amplifiers, and information transmitters \cite{mirbabaie2020social}. Other studies rather highlighted the impact of content characteristics such as emotionality \cite{stieglitz2013emotions} or the attachment of images and videos \cite{jung2018information}. Notably, most previous studies focused on a single platform such as Twitter, thereby neglecting the reality of information diffusion, as is also the case for networked framing. In this regard, \cite{kane2014s} already argued that social media users can share content from Facebook to Twitter and vice versa, pointing toward the vanishing boundaries between different social media platforms. A study proposed that sharing viral videos on alternative platforms might affect their popularity on the original platform \cite{krijestorac2020cross}. Furthermore, scholars found a significant influence of mass media and external websites on information diffusion within a social media platform \cite{myers2012information}. This phenomenon is also known as the spill-over effect among communication scientists \cite{mathes1991role}. The detection of spill-over effects represents a methodological challenge as it is difficult to find the origin of information in the online media sphere \cite{pfetsch2013critical}. The authors suggested using a crawler to identify URLs linking to other online information sources. Jung et al. \cite{jung2018information} built upon this methodological approach and demonstrated in a case study that online news sites referenced more frequently to information from Twitter than vice versa. With our study, we aim to extend the scarce research body on multi-platform networked framing and cross-platform diffusion by examining the occurrence and diffusion of frames within and across social media and online media.

\section{Research Design}
\label{sec:3}
In this section, we explain the steps involved in performing the study. They are discussed below.

\subsection{Data Collection}

In this section, we explain the data collection from different platforms. For a holistic picture of the online discourse, Twitter and YouTube, as well as Norwegian newspapers, were chosen. While Twitter is known for (textual) breaking news content, YouTube is the most prominent platform for video content. Therefore, these two social networks were considered especially useful for this study. The time frame for the data collection was 12th November 2019 until 30th November 2019. A detailed explanation of each platform is given below. For all platforms, we decided that the keyword \textit{SIAN}, as the organizer of the event, such as the keywords \textit{Arne Tumyr} and \textit{Lars Thorsen}, as the involved SIAN leaders, were relevant. Furthermore, the keyword koranbrenning was chosen as it was the most prevalent term to describe the event in the Norwegian news. Finally \textit{Kristiansand} was chosen as a keyword, as the incident’s location. The results of the data collection confirmed that the keywords delivered relevant results, wherefore the keyword selection was not further adjusted.

\subsubsection{Twitter Data}

The Twitter data was gathered with a self-developed Python crawler, which connects to the Twitter Search Application Programming Interface (API) before the commercialization of Twitter data) and collects tweets using keywords SIAN, Arne Tumyr, Lars Thorsen, koranbrenning. The tool collected all Norwegian tweets, retweets, and replies that contained at least one of these keywords and were published from 12th November 2019 until 30th November 2019. We manually checked the search results for relevance and excluded 57 tweets that were not related to the Quran burning in Kristiansand, excluding tweets about another Quran burning that happened in Sweden or tweets about other activities of SIAN, Arne Tumyr, and Lars Thorsen. The final dataset included 2,267 tweets consisting of 865 original tweets, 1,131 retweets, 224 replies, and 47 commented retweets. The 1,131 retweets were excluded from analysis and treated as duplicates as they did not  add new frames to the Twitter discourse. The final Twitter data set contained 1,136 tweets.

\subsubsection{News Paper Articles}

For tracking the news media, we used commercial software by M-brain (now Valona\footnote{https://valonaintelligence.com/}) to monitor media, ensuring comprehensive news coverage. It provided us with openly accessible content and those behind online paywalls. We used the search terms "koranbrenning" and "Kristiansand" to browse the content from the news media. M-brain searched all Norwegian media outlets and returned the articles which match the search terms. From the obtained results, we further filtered the news articles which contained any of the search terms "Lars Thorsen", "Arne Tumyr", or "SIAN". The final news media data set included 115 news articles that had been published between 12th November 2019 and 30th November 2019. After duplicates were deleted, 109 articles remained for the analysis. The data set included the article heading, URL to the news article, and date of publication, and we crawled the content of the news article. The media outlets can be categorized into three groups: State-owned and mainstream media, Online and Independent Media and Special Interest and Non-News Platforms. In addition, we also added to this classification who owned the media outlet and their affiliation to the Norwegian Press Organization, such additional notes if applicable. For a detailed overview please refer to table 2 in the Appendix.

\subsubsection{YouTube Videos and Comments}

For the YouTube analysis, we collected videos and comments related to the Quran Burning incident. To identify the relevant videos, we used the YouTube keyword search and conducted four different manual searches of the keywords “SIAN Norway,” “Arne Tumyr,” “Lars Thorsen,” and “Koranbrenning.” The relevancy was assessed by reading the title and description and watching the video. Some videos were excluded, for example, when the videos were related to a different Quran burning incident. Videos that occurred in multiple searches because they included several keywords were included only once. Overall, we excluded five duplicate videos. We identified 112 relevant videos in total. Out of them, 71 videos had comments. We crawled all 8,917 comments related to those videos. As the total number of comments was too extensive for a manual assessment, we applied a sampling approach to all videos with more than 100 comments. Ten videos had more than 100 comments, and 62 videos had less than 100 comments. For the sampling, we took all comments from the videos with less than 100 comments, and for videos that had more than 100 comments, we sampled 100 comments from each video. Overall, we got 2,041 comments from 71 videos in total. This approach provided us with a diverse data set of all comments from all videos. In contrast to the newspaper articles and tweets regarding the Collection of YouTube videos and comments, there was no language restriction to Norwegian and English, but all videos and comments have been analysed with help of the subtitles provided by YouTube and translation software whenever possible.

\subsection{Data Analysis and Preprocessing}

In this section, we describe the steps involved in the data preprocessing and analysis. We have applied a series of steps to clean our data set. We filtered the URLs from the text of tweets, YouTube comments, and news articles for data cleaning following the approach mentioned in \cite{shahi2020fakecovid}. After that, we manually identified the domain of the URLs to identify the link target platforms (i.e., Twitter, YouTube, online news sources, and others). The category others included links leading to other social media and online news sources that we have not included in our sample (e.g., Facebook, Instagram, religious websites).

\subsection{Frame Analysis}

As the Quran Burning incident in Kristiansand can be classified as a politically motivated event, we decided to use an existing codebook for the analysis of policy frames by \cite{boydstun2014tracking,card2015media}.
The codebook has been developed and validated with a pilot study covering three major events in the US. We used the most recent version of the codebook, which was updated in 2016. The codebook was developed for the analysis of policy debates in the United States and consists of 15 frame dimensions. The 15 frame dimensions have been adapted according to the Norwegian context and the context of the Quran Burning incident. Each dimension was equipped with a short paragraph about the relevance of the Quran Burning case, possible keywords, and examples, which can be found in Table 1 in Appendix. As the data from Twitter, YouTube and newspaper articles confronted us with different conditions the codebook was slightly adapted for each medium, which will be explained along the description of the coding process. As a first step of the framing analysis, the coders intensively studied the codebook and then added relevant paragraphs, keywords, and examples. In the next step the coders have been provided with the data sets and their English translations. After reading the text, the coders needed to decide if the translation was understandable or if there was further clarification needed. In case there was a better translation of Norwegian data required, the Norwegian team members were involved. If the translation was understandable, coders needed to decide if the information was relevant for the respective case study. To understand the YouTube videos, the subtitle function was used if available, and the title and description were translated with the help of Google Translate. Afterwards, the relevance of the data was evaluated. The first relevant criterion was whether the tweet, video or comment was about or in connection with the Quran burning in Kristiansand. If this was not the case, it was not deemed to be relevant. Exceptions were made if the covered international incident was a reaction to the Quran Burning in Kristiansand. Messages which only contained an emoticon or random strings of characters were marked as irrelevant. The tweets of the Twitter data set have been sorted according to the respective tweet ID in order to identify the conversational threads to which they belonged. Each tweet, YouTube video, YouTube comment, or newspaper article received one primary frame. The frame could vary from the original tweet in case of commented retweets and replies. In case of more complex messages, especially regarding newspaper articles a secondary frame was chosen if it was not possible to reduce the main messages of the text to one primary frame. Each data set was coded by two coders. First of all, 20\% of each data set was coded independently to check the intercoder reliability of each coding. The intercoder reliability was calculated in the form of Cohen’s Kappa coefficient which can be found in Table 4 of Appendix. 
If the result was satisfying, the coding process was continued. In case of deviations in the first coding round, a third coder was involved, and the majority rule has been applied. The rest of the data sets were divided between the first two coders.

\section{Result}
In this section, result obtained from different analysis is provided. 
\subsection{Framing Analysis}
With help of the code book we identified the different frames, which have been used in the discourse and media coverage about the Quran burning incident in Kristiansand. We have seen a distinct usage of frames in news articles, tweets and YouTube videos and comments. A detailed description of annotated frames is given in Table 3 in the Appendix. For the news articles, primary frames and secondary frames have been applied regularly, as the longer texts often did not allow being reduced to one single frame. For the  coding of the YouTube comments the frame dimension "None" was applied more frequently than in comparison to the other media types. By having a closer look at this frame we realized that this was the case because many of the comments could not be translated for analysis and the automated translation had reached its limits.
In the 109 coded news articles, the frame dimension Legality, Constitutionality \& Jurisdiction was most frequently used, with 50 (45.9 \%) references. That means that most of the news articles had legal issues as their main subject in the form of references to freedom of speech or constitutional issues. Very prevalent was the discussion if it was legal, according to the Norwegian constitution, to burn the Quran. One example of the use of the legality frame dimension is an article by the online news outlet document.no, with the title “It should not be allowed to burn holy books”. The article discussed if the decision by the Police Directorate to stop SIAN from burning a copy of the Quran, as a violation of section 185 of the Penal Code on the prohibition of hate speech, was well grounded. The second most applied frame dimension in the news article data set was the External Regulation and Reputation frame with 12\%. The incident in Kristiansand led to direct reactions in the Muslim world, especially Pakistan. The articles in which this frame was used often discussed the implications of the Quran burning for the Norwegian telephone provider Telenor, which owns more than a quarter of the Pakistani cellular market [38]. The public service broadcaster NRK published, for example, an article with the title “Call for boycott of Norwegian companies after Quran burning”. The article described the reactions in the Pakistani community and media landscape to boycott Norwegian products and especially the Pakistani branch of the Norwegian telephone provider Telenor. In addition to that, articles with the External Regulation and Reputation frame described the political reactions to the incident in the form of an appointment of the Norwegian ambassador in Pakistan for a statement on the case by Prime Minister Imran Khan and the Ministry of Foreign Affairs. An example for that is the article “Norway is called on the carpet” (the Norwegian proverb to call someone to the carpet, can be translated with “to scold someone”) by the Norwegian daily newspaper Dagbladet. It discussed that the Pakistani Ministry of Foreign Affairs scolded Norway’s ambassador and expressed concern that a Quran was set on fire in Kristiansand. The frame dimension Security and Defense was with 10.1\% also applied regularly in the news article data set. The articles, which contained this primary frame, discussed extensively that the Quran burning incident resulted in a concrete security threat for the Norwegian state. One outlet which used this frame is e.. the news site resett.no, which is a controversial online news site in Norway, known for its skepticism regarding immigration and Islam. Resett.no is not part of the Norwegian Press Organization due to the fact that the organization has deemed that they do not adhere to the Norwegian Press Codex. The article of resett.no using the Security and Defense frame has the title: “Threatened to kill after Quran burning: Kill him please!” It discussed that the burning of the Quran has created violent reactions both on social media, but also from Muslim communities in Norway. The article stated that several Muslim leaders in Agder warned that they will now report SIAN to the police. Besides that, the Security and Defense frame was applied in articles, which reported about protest actions, which were triggered by the incident. One example for that is the article “The Norwegian flag is burned in protest against the burning of the Quran ” by the local newspaper Fedrelandsvennen, which reported that the Norwegian flag was burned during demonstrations in different Muslim countries among others in the state of Karachi in Pakistan. The authors interpreted these incidents as a sign of a concrete threat against the Norwegian state to suffer an Islamist-motivated attack. Due to the complexity of the newspaper articles in several cases a secondary frame was chosen. The most common secondary frame was again Legality, Constitutionality \& Jurisdiction with (32\%), which underlines the importance of the legality frame in the news article data set. Besides that as well the Morality and Ethics (18\%), Political Factors and Implications (16\%)\, and Security and Defense (14\%)\ frames played an important role as secondary frames. The article “Koran burning, blasphemy, freedom of speech and hate speech” is an example of an article with a primary and a secondary frame. The primary frame is Legality, Constitutionality \& Jurisdiction and the secondary frame is Morality and Ethics. The article discusses both the legal interpretation of the incident and its morality, following the stance that not everything which is legal is as well morally defensible.

\subsection{Tweets}
Of the 1,136 analyzed tweets, 47 tweets were coded as irrelevant, which are  4\% of the data set. Among the relevant tweets, the Morality and Ethics frame was with 28.1\% the top frame. An anchor example for this category is: “I am not a true follower, but what if the Bible was burned? Would antifa violence responders react? I doubt.”. In addition to the Morality and Ethics frame the Political Factors and Implications frame dimension was used in 19.2\% of the sample. There have been a lot of comments after the event which underlined a political motivation. Both lay people and politicians used the burning of the Quran to voice political views, coming from different political camps. Several accounts pointed out that they have the impression that SIAN received more attention, that it deserved according to their overall importance for the Norwegian political landscape. One anchor example for that is: “Given that the only Norwegian party of significance that is close to SIAN has just made a historically poor municipal election, SIAN should not imagine that they have any significant support.” The accounts, who advocated for SIAN often pointed out that SIAN is a relevant political group. In the following example SIAN is presented as a moderate, non violent group in comparison to political groups of the left political spectrum: “Then you should stop using the word extremist about SIAN. They have never resorted to violence, which your dear Communists in Red and Antifa constantly do”. Besides that, only the Legality frame reached a relevant number of tweets with 13.6\%. Two examples for the usage of the Legality, Constitutionality \& Jurisdiction frame are:\\
“Penal Code \textsection185 used against burning of the Quran? Seems like the law stretches quite a bit.” and “Norwegian scandal: Freedom of speech is something we play. The police had received a secret illegal order to stop "violation of the Qur’an" when the SIAN (Stand Islamization of Norway) demonstration in Kristiansand was brutally interrupted by local police.” For annoated tweets without none frame, around 16\% (191 tweets) of all tweets referred to external media sources. 55 of those tweets have been published by media organizations themselves, the others were shared by mainly individual accounts and some organizations. As these tweets neutrally shared media links, with short article snippets instead the content of the shared media content was coded for the analysis. All other frames covered only in between 0.2\% and 7.3\% of the data set and are thus not described in more detail. A second frame added to less than 5\% of the tweets and can thus be ignored.

\subsection{YouTube videos}

All YouTube videos that only showed the incident as a whole without commenting on it were coded as None if there was no frame included either in the title or description of the video. The Morality and Ethics frame was the most prevalent frame used in the YouTube videos, with 36.6\%. However, a minor number of videos applied to the External Regulations \& Reputation frame, which together account for 9.9\% of the data set. 


\begin{figure}
    \centering
    \begin{minipage}{0.33\textwidth}
        \centering
        \includegraphics[width=0.89\textwidth]{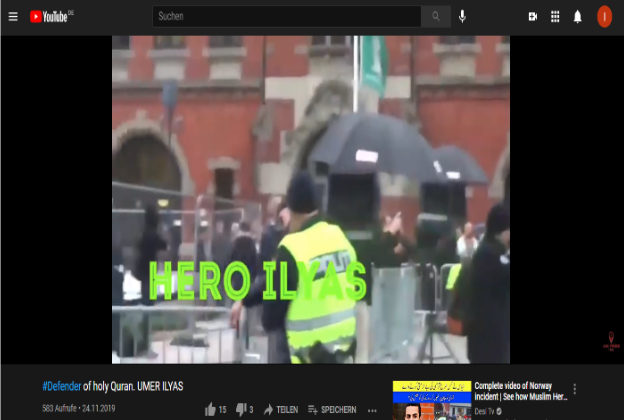} 
    \end{minipage}\hfill
    \begin{minipage}{0.33\textwidth}
        \centering
        \includegraphics[width=0.99\textwidth]{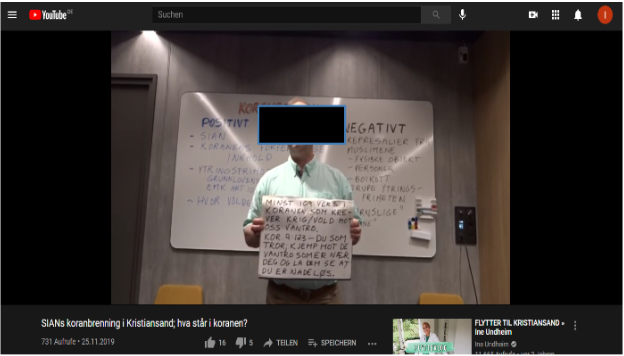} 
    \end{minipage}
     \begin{minipage}{0.33\textwidth}
        \centering
        \includegraphics[width=0.99\textwidth]{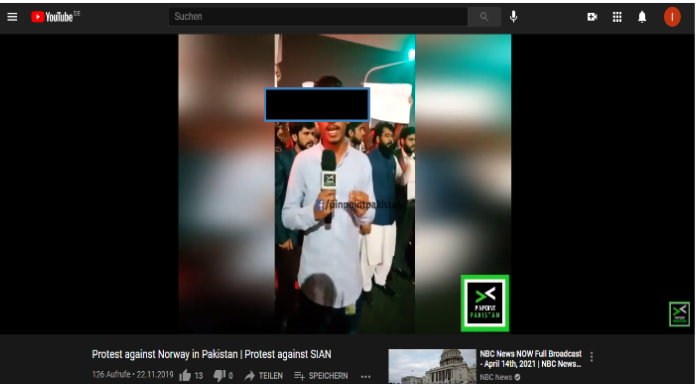} 
    \end{minipage}
    \label{fig:parameter}
    \caption[]{Annotated YouTube videos(From left to right video a,b,c)
}
\end{figure}

The video in Figure 1 (a) was uploaded to YouTube by channel JHUNJHUNU PRIME TIME, joined YouTube on the 7th of November 2019, only a few days prior to the incident. Even if the account is supposed to give the impression of an official station due to the name and the chosen avatar, it can be assumed that it is not an official source but a channel run by a private person. Another indicator for that is that the channel description is incomplete. Another example for the use of the morality frame is (b) in Figure 1 a video of the leader of the Norwegian PEGIDA (Patriotic Europeans against the Islamization of the Occident). In his video with the title “SIAN’s Koran burning in Kristiansand; what does the quran say?“, which he prepared in reaction to the SIAN burning of the Quran in Kristiansand, he presents his interpretation of the five pillars of the Quran. This video is a good anchor example for videos which have been created by sympathizers of SIAN. The videos, which belonged to the External Regulation and Reputation frame, often showed and described protests, which came up in Muslim countries, especially Pakistan, after the incident. An anchor example can be the following video. This video (c) in Figure 1 was posted by the account Pinpoint Pakistan, which has a relatively professional appearance, such as JHUNJHUNU PRIME TIME. The channel contains more information about its scope and gives a contact email for questions. However, there is only a YouTube and Facebook Page with the name of this organization and no official media house under this name, which as well gives the impression that it is a media channel run by private persons. It needs to be doubted that the video thus has been produced by the channel owners themselves. 


\subsection{YouTube comments} 

Coding the YouTube comments, we used fewer frames than the other channels, pointing as well at the limitations of using a predefined codebook. The frames that were most used were Morality and Ethics, None and Other. The reason why Morality and Ethics was the most frequently used, was the reference to religion, e.g.: “Long live Allah.” Often the morality frame was also used to express beliefs in a hateful manner: “I pissed in koran and ur momz!”. In the frame Other we have put comments that did not fit in any other category, but were still relevant. Often there was shortly formulated support or dislike formulated in a very short manner like “Good” or “Nice” or “lionhearted” or from the SIAN supporters: “Thanks Lars”.  Some have also been hate speech like: “Look at the red pig he looks like”. The reason why so many comments were coded as none, was because the content either did not make sense, or we did not understand the meaning of it. While some meaning might have been lost in translation, there have been comments including random combinations of letters. 
We also saw 5.2\% of comments about Norway’s reputation, so we coded these External Regulation and Reputation. Examples are “Fuck Norway, boycott Norway”, “Norwegians are bastards” and “Shame on Norway”. This could have harmed Norway’s reputation and were often as well in a hate speech manner.



\subsection{URL Analysis}
We coded 888 tweets, 27 YouTube comments, and 68 news articles during the manual framing analysis, including reference hyperlinks from one of the other platforms. However, some tweets, comments, and articles included multiple URLs. Summing up, those that included more than one URL led to the total number of 988 URLs from tweets, 27 from YouTube comments, and 456 URLs from news articles from the above spillover links. If there were multiple URLs from one tweet, YouTube comment, or newspaper article to another platform, we split it into different URLs, so only a link is associated with each source and the target node. We further analyzed the domain of the URLs to get information about the spillover of hyperlinks. To identify cross-platform diffusion and spillover effects, we categorize the URLs into four categories based on domains: Twitter, YouTube, News Media, and others (which include other domains apart from the above three, like social media platforms Facebook and Instagram.). Figure 2 describes the number of different URLs shared from one platform to another. The figure shows the diffusion of URLs from each platform to four different categories: Twitter to news media, YouTube, Twitter and others. Overall, news articles contain multiple URLS within the articles, mainly referring to other news articles. On Twitter the tweet authors as well use references within the own newtork boundaries by referencing tweets, but as well share URLS of news media and YouTube. In contrast YouTube contains limited URLS, mainly staying within own network boundaries.

Within our collected datasets, we analysed the spillover of URLs among news articles, Twitter and YouTube. We matched URLs of news articles to Twitter and YouTube. A spillover of URLs from news article to another is observed mainly on Twitter; around 26\% unique news articles are posted 171 times on Twitter from domains such as \textit{utrop.no, resett.no, gjenstridig.no, nrk.no, dagladet.no, document.no, afternposten,vg.no, rights.no, nrk.no}. At the same time, only two news articles are shared on YouTube comments, which are not explicitly from our datasets. For cross-platform, 17 YouTube videos are shared on Twitter. In contrast, among YouTube comments, only a few URLs are mentioned in comments, mainly on other YouTube videos and some other social media platforms such as Facebook and VK.

\begin{figure}
    \centering
        \includegraphics[width=0.89\textwidth]{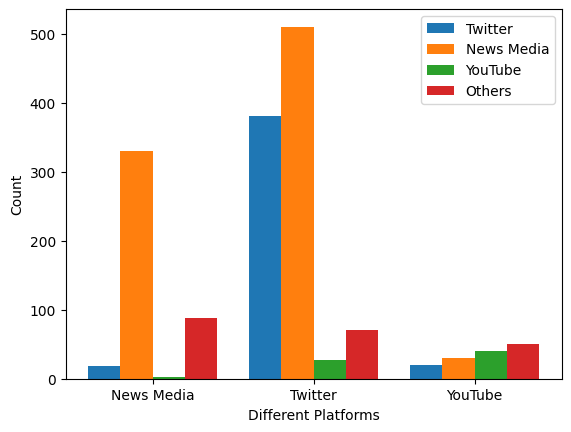} 
    \label{url}
    \caption{Distribution of URLs and their source on three different platforms.}
\end{figure}

\section{Discussion}

This study offers insights about two main topics: differences of multi platform framing of a social incident on social media (Twitter and YouTube) and online news sites (RQ1) and cross-platform diffusion of topics and frames (RQ2). The content analysis of the online news site articles, tweets, and YouTube videos and comments resulted in a distinctive application of frame dimensions on the different platforms. Of the 14 main frame dimensions adapted from Boydstun et al. \cite{boydstun2014tracking}, only four frame dimensions reached a threshold of at least 10\% in one of the data sets, namely Legality, Morality, External relations, Political factors and the two dimensions None and Other. While the legality frame dominated the discourse on online news sites, the morality frame dominated the two social media data sets. We argue that the high dominance of the morality frame in the social media data sets is related to the higher level of personal communication on social media platforms, which allows each individual and organization to contribute to the discussion \cite{bruns2018gatewatching}. According to \cite{stahl2012morality} moral intuition forms the very basis of any normative evaluation \cite{stahl2012morality}. It is an unfiltered and non-reflected reaction if an incident should be classified as right or wrong \cite{kekes1986moral}. We argue that the use of the morality and ethics frame might be more prevalent in social media as individual users are more likely to express their moral intuition than (professional) journalists. Further, we considered international sources in the YouTube data set too, wherefore here we also see reactions from people living in Muslim countries like Pakistan, feeling morally offended. The harshest formulation of personal opinions was present in the analyzed YouTube comments, which involved only the frame dimensions of morality and other. The other category was used particularly frequently for YouTube comments, including hate speech which could not be assigned to any of the existing frame dimensions. Furthermore, YouTube comments showed the highest level of polarization, particularly within the morality frame. This polarization is reflected by potential echo chambers of the Muslim supporters and right-wing supporters, predominantly encountering viewpoints that reinforce their existing beliefs, which might intensify polarization. The dominance of the morality frame, which provokes strong emotional responses, contributes to this phenomenon. Polarization can have broader societal impacts, including increased social tensions and greater distance between different groups or countries. The problem of misinformation, which is linked to highly polarized discourse environments, extends beyond false information to the intensive diffusion of specific frames and narratives that shape public beliefs and perceptions \cite{Starbird2024}. According to Starbird, these frames influence the evidence used in sensemaking processes and guide interpretations and the formation of public opinion. Our results suggest that the frames and narratives surrounding the Quran burning event may overshadow existing evidence, exacerbating societal and political polarization. However, the analysis of the Twitter comments revealed as well the limitations regarding the application of Boydstun’s codebook on YouTube comments. In further research, a more specific codebook should be developed for this medium and discourse space.
The fact that that polarization was less prevalent on Twitter could potentially be explained by the different user groups of the platforms, or differently strict or successful countermeasures of the platforms regarding hate speech \cite{alkiviadou2019hate}. The deletion of hate speech should be practiced more strongly by YouTube. Not only at the video level but, above all, in the comments, where AI-supported approaches may be used. The perspective in the YouTube comments was very unified and contained mainly narratives like the attackers of SIAN are heroes, Islam is a religion that should be praised, and Norway should be reproached for not respecting and protecting Islam. We argue that due to the language barriers, as most of these comments have been originally published in Urdu or Arabic, there was little variation and counterspeech of different opinion camps. Research has shown that users do interact barely with textual posts in foreign languages \cite{lim2017making}. While passive interactions in the form of likes happen, commenting takes place very rarely [43]. Therefore, we assume that Norwegian or English-speaking YouTube users have not engaged and reacted with the comments in Arabic or Urdu, which explains the non-existent counter speech. It also explains a different application and diffusion of frames within different user/language groups and shows that the idea of networked publics and the networked framing has certain boundaries \cite{meraz2013networked}. However, the traditional media act as a bridge actor here, presenting the incidents of hatred from social media in other countries such as official reactions from countries such as Pakistan (External relations frame), which underlines the role of traditional news media in cross platform frame diffusion. 
In contrast to YouTube and the online news sources the discourse on Twitter was the most diverse according to the variety of applied frames, although the threshold of 10\% was not exceeded for all frame dimensions. It indicates that Twitter might represent a more diverse discourse space compared to the other platforms. However, this should always be considered with the caveat that the topic itself may also have influenced how it was discussed on which platform. For further generalizations future studies are indispensable. Based on the Kristiansand incident, we perceived it as unexpected that meta-discourses under the frames of Fairness and Equality or Cultural Identity did not play a serious role in the data set. There was also no difference in the online news sources, which offer more space for societal discourses. Instead of an overall societal discussion about the role of Islam and Muslim groups in Norwegian society, the discussion was more focused on the personal moral classification of the conflict, or in online news sources on the legality of the committed act. This points to a potential question for future research to which extent social networks allow and foster meta-discourses or whether these are lost in the mass of personal statements. After analyzing the distribution of the different frames among the three analyzed platforms (Twitter, YouTube and online news sites), the analysis of URLs used in the data set was implemented to answer RQ2. First, we categorized the links originating from a platform according to their platform type: Twitter, YouTube, news articles, and others. The category others included all links that did not fit the three main categories and included i.e. references to Instagram, Facebook, and religious websites (e.g. islam.net). We observed that a significant amount of URLs diffusion is happening within the same platform for news articles and YouTube comments. While around 67\% of all URLs in YouTube comments point to other YouTube content, 72\% of all online news sources point to other online news sources while only 39\% of URLs mentioned in tweets refer to other tweets. Around 47\% of URLs mentioned in collected news articles refer to the same media house, which indicates that online news sources are more likely to share sources from their own company to strengthen their own economic goals of longer dwell time and clicks. Solely on Twitter, 52\% URLs mentioned in tweets referes to news media, 3\% to YouTube, 39\% to other tweets and 8\% to other platforms .
This strengthens the ideas of \cite{myers2012information,wang2020understanding}, who identified news media as influential drivers of diffusion in social networks. In contrast, the URLs of news sites in the Quran Burning case study only included roughly 8\% of the URLs pointing to different social media sites, including Twitter and Facebook. This contradicts the findings of Jung (2018) \cite{jung2018information}, who found out that news media were referencing Twitter more often than Twitter posts referred to news media. However, Jung (2018) \cite{jung2018information} did not use a data set looking at one specific case study but a broader variation of topics. Although the diffusion and referencing within the respective platform boundaries are high, cross-platform diffusion makes up 25-60\% of the sub datasets of this case study, strengthening the results of \cite{kane2014s} and \cite{krijestorac2020cross}, who both found an interdependence between the degree of diffusion and cross-platform spill-overs. While for the spill over of news articles to Twitter it could be identified by our coding, that in many cases shared news articles also led to a share of frames, as snippets of the articles formed as well part of the tweet, it remains unclear is if the other reference spill-overs which have been identified in the data sets do also directly lead to a frame spill-over, which could give insights into the importance of the different platforms and actors involved in the frame-setting and agenda-building process \cite{pfetsch2013critical}.

\section{Limitations and Future Research}
\label{sec:6}
One major limitation of our keyword-based data collection approach is that some relevant data about the Quran burning incident in Kristiansand might have been missed, compromising completeness. Furthermore, the fact that an existing codebook was adapted, may be another methodical limitation. A codebook derived from the data set always has the strength of being more case-related and specific. While the Twitter data and YouTube videos could be easily coded with the media corpus codebook, 
We see the limitations of the codebook for YouTube comments and understand an adoption of it for YouTube comments as a potential future research endeavor. We admit that there is a certain language bias in relation to the collection of YouTube videos and comments in our analysis. However, we argue that videos are more accessible across languages compared to text alone, and AI-generated subtitles on YouTube were sufficiently accurate for our analysis. In addition to the limitations the study also stimulates many new ideas for further research. First, in a follow-up study, it would make sense to code not only the dataset collected in the first step, but also all the third-party content identified by the URL analysis on the three platforms. Through this snowball system, it would be possible to make clearer statements about frame spill-overs in the future. This further analysis could for example be supported by a network analysis to visualize the spill-overs in the different directions and could help to identify the main frame-setters. Furthermore, it would be interesting to investigate whether sockpuppeting or astroturfing took place to push certain frames (e.g. by following the approach presented in [44]. Furthermore, it would also be appropriate to focus on the questions raised in the discussion of this article on the topic of discourse diversity on different platforms and the possibility of meta-discourses in social networks.

\section{Conclusion}

This study contributes to closing the research gap of networked framing on multiple platforms by analyzing how the socially and politically relevant incident of the Quran burning by the Anti-Islamic group SIAN in Kristiansand was framed on Twitter, YouTube and in online news sources. The analysis revealed that on social media the frame dimension of morality and ethics was highly dominant, while on online news sites the legality frame played the most relevant role. The higher use of the morality frame on social networks can be related to two main reasons: the higher number of personal statements in social networks and the involvement of users from Islamic states like Pakistan who were morally concerned by the incident. Another finding was the high appearance of hate speech in the form of the other frame dimension in the YouTube comments, which was not similarly present on other platforms. As most hate speech was published in Urdu and Arabic there was no counter-speech in English or Norwegian, underlining a lack of interaction between different language groups and ergo a limited diffusion of frames between different language groups. Which underlines the limits of the proposed concept of networked framing and gatekeeping by \cite{meraz2013networked}. Although the internationality of the social media users was higher on YouTube, the frame dimension of External regulations and \& Reputation was more relevant in online news sources. However, online news sources were highly dominated by the legality frame, discussing the legal classification of the incident in accordance with the Norwegian law, which did only play a minor role on social networking sites. The highest diversity of frame dimensions was applied by the Twitter community pointing at a more manifold discussion of the topic on this platform. However, meta-discourses on the fairness of the treatment of minority groups or a discussion on the cultural identity of Norwegian society in relation to the incident have been absent. Concerning the diffusion of content regarding the Quran Burning incident between different social media platforms and online news sites, it could be revealed that most diffusion takes place within the specific network boundaries. On Twitter, the highest amount of content generated outside the platform boundaries was shared. The most self-contained platform in this study was YouTube, which leads to the assumption that, at least for this case study, YouTube has a limited networked public solely looking at the content created by its own community. Cross-platform diffusion forms between a quarter to two thirds of the disseminated content, which underlines the interconnectedness and influence between the different discourse spaces and platforms. The study underscores the discursive power of social network users, who do not solely copy the frames of traditional media one-to-one but apply different and more multifaceted frames in their reflections. The greatest linkage was found between Twitter and online news sources for which we could identify not only a reference- but as well frame spill-over, as to a great extent frames of the news sources were spilled in form of text snippets and uncommented URLS into the Twitter community.

\section*{Acknowledgements}
The author(s) disclosed receipt of the following financial support for the research, authorship, and/or publication of this article: This project has received funding from the European Union’s Horizon 2020 research and innovation programme under the Marie Skłodowska-Curie grant agreement No 823866.

\section{Appendix}



\section*{Supplementary material}

# Appendix

**Table 1**: Summary of the original codebook and the described frame cues by Boydstun et al. [13] and the case specific additions to the codebook including keywords and phrases

| Frame dimension | Relevance for case study | Relevant keywords and phrases |
| --- | --- | --- |
| Economic | The costs, benefits, or any financial implications of the issue (to an individual, family, organization, community or to the economy as a whole). Impact on Norwegian companies worldwide. E.g. Telenor, one of the major mobile companies in Pakistan, faced reputational damage and customer loss. | Telenor, Pakistan, #boycottNorge |
| Capacity & Resources | The lack of—or availability of—time, physical, geographical, space, human, and financial resources, or the capacity of existing systems and resources to implement or carry out policy goals. Discussions about limited HR resources of the police to prevent the escalation during the Quran Burning incident. Discussions about the Quran Burning incident focusing on inability of the government due to | lack of police, lack of money, lack of investment |

| | | |
|---|---|---|
| | missing financial or human resources to monitor extremist movements. | |
| Morality & Ethics | Any perspective that is compelled by religious doctrine or interpretation, duty, honor, righteousness or any other sense of ethics or social or personal responsibility (religious or secular). As the Quran is a religious book and Muslims are a religious community, there are several articles, which contain morality frame cues. Furthermore, the incident triggered discussions about freedom of speech vs. hate speech, which can both be a morality or legality discussion. As well nationalist narratives that the Western culture needs to be protected etc. might fall under this frame. | Religion, Quran, Allah, Freedom of Speech |
| Fairness & Equality | The fairness, equality or inequality with which laws, punishment, rewards, and resources are applied or distributed among individuals or groups. Also the balance between the rights or interests of one individual or group compared to | Discrimination in relation to religion, gender, sexual orientation etc. |

| | | |
|---|---|---|
| | another individual or group<br>As the Muslim community is a minority group in the Norwegian society the fairness and equality frame might apply. | |
| Legality, Constitutionality & Jurisdiction | The legal, constitutional, or jurisdictional aspects of an issue, where legal aspects include existing laws and court cases; constitutional aspects include all discussion of constitutional interpretation and/or potential revisions;  and jurisdiction includes any discussion of which government body should be in charge of a policy decision and/or the appropriate scope of a body's policy reach<br>Discussions on the legality of the burning of the Quran. Is it forbidden by the constitution? As well the legality to generally burn something on a market square was discussed heavily. Furthermore, the Penal Code paragraph 185 (Hate Speech paragraph), was highly discussed in relation to the Quran Burning incident. | Hate Speech paragraph, (Blasphemy) paragraph, Penal Code |

| | | |
|---|---|---|
| Crime & Punishment (Retribution) | The violation of policies in practice and the consequences—retribution—of these violations. As a consequence of the event, seven people were indicted. Three for nuisance, two for violence against the police, and two investigation cases. Even though the incident was potentially legal, it caused law-breaking actions. | punishment, fine, conviction, crime |
| Security & Defense | Any threat to a person, group, or nation, or any defense that needs to be taken to avoid that threat. There have been calls for revenge. The Norwegian flag was burned in Karachi, Pakistan, as a protest. Threats towards Norway have been announced, and people were worried about potential terrorist responses to the incident. | revenge, terrorism, terrorist threat |
| Health & Safety | The potential health and safety issues. During the protests some people got attacked. If they were physically attacked they felt that their safety and health was at risk. | beaten, hit |
| Quality of Life | The benefits and costs of any policy on quality of life. | fear of revenge and violence |

|  |  |  |
|---|---|---|
|  | The incident has caused fear for revenge. Living in fear and being afraid of new violent situations to occur can have an impact on quality of life. |  |
| Cultural Identity | The social norms, trends, values and customs constituting any culture(s), as they relate to a specific policy issue. Discussions about the struggles of muslim immigrants to find their place in Norway and constant reminders that Norway is not their country of origin and home. Racist and nationalist narratives that people from different countries and cultural backgrounds should not mingle to protect the Western culture and the Western value system. | diaspora, not feeling at home, feeling unaccepted |
| Public Sentiment | The public's opinion. Includes references to general social attitudes, protests, polling and demographic information, as well as implied or actual consequences of diverging from or "getting ahead of" public opinion or polls. The public sentiment plays a role regarding the Quran | Public opinion polls revealed, The majority of the Norwegians |

| | | |
|---|---|---|
| | burning incident e.g. in the form of discussions about the public impression that Norway might not be safe anymore. But as well all stories about certain interest groups (nationalist groups and left-wing groups, minority groups or organizations) can deliver information about the public sentiment in Norway during that time. | |
| Political Factors & Implications | Any political considerations surrounding an issue. There have been a lot of comments after the event which underlined a political motivation. The right-wing side argued for the idea of Muslims not respecting the values of a western society. The left argued that these kinds of provocations and the right-wings reaction to the situation was done to promote the idea of Muslims being a problem and a threat to society. Both people and politicians used the burning of the Quran to voice political views. | The political party Fremskrittspartiet (FRP) and any other party in Norway and elsewhere, names of politicians |

| | | |
|---|---|---|
| Policy Description, Prescription & Evaluation | Existing policies, policies proposed for addressing an identified problem, as well as analysis of whether hypothetical policies will work, or existing policies are effective. There are stories, which discuss if the current political strategies on how to integrate immigrants into the Norwegian society are working. | immigration law, mandatory language and integration courses |
| External Regulation & Reputation | In general, Norway's external relations with another nation; the external relations of one Norwegian state with another. The event has received international attention. It led to a press release by e.g. the Turkish government or Pakistani politicians. | Ambassador / Ambassadors names, international relations, reputation of Norway |
| Other | Any frame cue that does not fit in the first 14 dimensions. | Example: "AMAZING GOOD HUMOR, GOOD !!!!!!!! What was the theme here? Quran Burning?" |
| None | Short messages that did not contain any form of framing, but solely a objective description of the situation | Messages like: The Quran was burned on the market place of Kristiansand by members of SIAN. |

**Table 2**: Overview of online news media part of the online news media data set classified by ownership, affiliation and additional notes regarding the scope of the medium

| State-Owned and Mainstream Media | Ownership | Affiliations | Notes |
|---|---|---|---|
| **NRK (www.nrk.no)** | State-owned, under the Ministry of Culture | Member of the Norwegian Press Organization | Largest media house in Norway, non-commercial and politically independent. |
| **VG (www.vg.no)**, | Owned by Schibsted | Member of the Norwegian Press Organization | Largest newspaper in Norway, significant historical circulation. |
| **Dagbladet (www.dagbladet. no)** | Owned by Aller Media AS | Member of the Norwegian Press Organization | Daily newspaper with a focus on literature and cultural issues. |
| **TV2 (www.tv2.no)**: | Privately owned | Member of the Norwegian Press Organization | Major broadcaster with significant news coverage. |
| Online and Independent Media | | | |
| **Filter Nyheter (filternyheter.no)**: | Independent | Member of the Norwegian Press Organization | Focus on investigative journalism and long-read articles. |

| | | | |
|---|---|---|---|
| **Utrop (www.utrop.no):** | Independent | Member of the Norwegian Press Organization | First newspaper and TV for multicultural news in Norway. |
| **Fritanke (fritanke.no):** | Independent, associated with the Human-Ethical Association | Adheres to the editor's poster | Focuses on humanist and ethical issues. |
| **Document.no (www.document. no):** | Independent | Not a member of the Norwegian Press Organization | Conservative focus, covering major political and social issues. |
| **Resett (resett.no):** | Independent | Not a member of the Norwegian Press Organization | Right-wing, focuses on alternative news and opinions. The site has been taken down and is continued under a different name: inyheter.no |
| **Rights.no (www.rights.no):** | Human Rights Service (HRS) | Not a member of the Norwegian Press Organization | Focuses on immigration and integration issues, controversial for its stance on Islam. |
| **Gjenstridig.no (www.gjenstridig. no):** | Supported by Fritt Ord | Not a member of the Norwegian Press Organization | National conservative, critical of mainstream media. |
| **Lykten.no (www.lykten.no):** | Independent | Not a member of the Norwegian Press Organization | Focuses on challenging established media narratives, right-wing orientation. |

| | | | |
|---|---|---|---|
| **KSU.no (www.ksu.no)** | Independent | Not a member of the Norwegian Press Organization | Focuses on local news in Kristiansund and Nordmøre. |
| **Ekte Nyheter (ektenyheter.no)**: | Independent | Not a member of the Norwegian Press Organization | Minimalist state, libertarian focus, skeptical of mainstream media. |

Special Interest and Non-News Platforms

| | | | |
|---|---|---|---|
| **Imamen.no (imamen.no)**: | Ahmadiyya Muslim Community | Not a member of the Norwegian Press Organization | Platform for Ahmadiyya imams to present views and engage with the public. |
| **SIAN (www.sian.no)**: | Stop Islamization of Norway | Not a member of the Norwegian Press Organization | Opposes Islamization, promotes dissemination of information about Islam. |
| **Oppror.net (www.opprop.net )**: | Independent | Not a news outlet | Platform for creating and promoting signature campaigns. |

**Table 3**: Usage of primary frames and secondary frames in online news sources, tweets, YouTube videos, and YouTube comments. Total numbers in the URL analysis might differ, due to more than one URL per tweet, post or article.

| Frame dimension | online news sources | | tweets | YouTube videos | YouTube comments |
|---|---|---|---|---|---|
| | Count Primary | Count Secondary | Count Primary | Count Primary | Count Primary |
| Economic | 2 (1.8%) | 4 (8.0%) | 41 (3.6%) | 1 (1.4%) | 0 (0.0%) |
| Capacity & Resources | 0 (0.0%) | 0 (0.0%) | 2 (0.2%) | 0 (0.0%) | 0 (0.0%) |
| Morality & Ethics | 6 (5.5%) | 9 (18.0%) | 319 (28.1%) | 26 (36.6%) | 332 (16.3%) |
| Fairness & Equality | 1 (0.9%) | 0 (0.0%) | 28 (2.5%) | 1 (1.4%) | 0 (0.0%) |
| Legality, Constitutionality & Jurisdiction | 50 (45.9%) | 16 (32.0%) | 154 (13.6%) | 1 (1.4%) | 22 (1.1%) |
| Crime & Punishment | 3 (2.8%) | 3 (6.0%) | 28 (2.5%) | 2 (2.8%) | 0 (0.0%) |
| Security & Defense | 11 (10.1%) | 7 (14.0%) | 55 (4.8%) | 0 (0.0%) | 1 (0.0%) |
| Health & Safety | 1 (0.9%) | 0 (0.0%) | 42 (3.7%) | 0 (0.0%) | 0 (0.0%) |
| Quality of Life | 1 (0.9%) | 0 (0.0%) | 23 (2.0%) | 1 (1.4%) | 0 (0.0%) |
| Cultural Identity | 1 (0.9%) | 0 (0.0%) | 48 (4.2%) | 0 (0.0%) | 0 (0.0%) |

| | | | | | |
|---|---|---|---|---|---|
| *Public Sentiment* | *5 (4.6%)* | *1 (2.0%)* | *37 (3.3%)* | *0 (0.0%)* | *2 (0.1%)* |
| *Political Factors & Implications* | *10 (9.2%)* | *8 (16.0%)* | *218 (19.2%)* | *1 (1.4%)* | *0 (0.0%)* |
| *Policy Description, Prescription & Evaluation* | *4 (3.7%)* | *0 (0.0%)* | *22 (1.9%)* | *0 (0.0%)* | *0 (0.0%)* |
| *External Regulation & Reputation* | *13 (11.9 %)* | *2 (4.0%)* | *83 (7.3%)* | *7 (9.9%)* | *106 (5.2%)* |
| *Other* | *1 (0.9%)* | *0 (0.0%)* | *3 (0.2%)* | *1 (1.4%)* | *334 (16.4%)* |
| *None* | *0 (0.0%)* | *0 (0.0%)* | *33 (2.9%)* | *30 (42.3%)* | *558 (27.4%)* |
| ***Sum*** | ***109 (100%)*** | ***50 (100%)*** | ***1136 (100%)*** | ***71 (100%)*** | ***2031 (100%)*** |

**Table 4**: Cohen's Kappa Coefficient per medium and its interpretation

| Medium | Cohen's Kappa Coefficient | Interpretation according to Landis and Koch (1977) |
|---|---|---|
| Twitter | 0.86 | Almost perfect |
| YouTube videos | 0.89 | Almost perfect |
| YouTube comments | 0.87 | Almost perfect |
| Newspaper articles | 0.87 | Almost perfect |



\bibliographystyle{splncs04}
\bibliography{hate}

\end{document}